\documentclass[useAMS,usenatbib]{mn2e}
\usepackage{graphicx}
\usepackage{amsmath}
\usepackage{tabularx}

\usepackage{times}
\usepackage{amssymb,amsmath}

\def\lsim{ \lower .75ex\hbox{$\sim$} \llap{\raise .27ex \hbox{$<$}} }
\def\gsim{ \lower .75ex \hbox{$\sim$} \llap{\raise .27ex \hbox{$>$}} }

\newcommand{\fig}[1]{Fig.~\ref{fig:#1}}

\def\comp{\,c/\omega_{\rm p}}
\newcommand{\bi}{\begin{itemize}}
\newcommand{\ei}{\end{itemize}}

\def\comp{\,c/\omega_{\rm pi}}

\title[X-ray polarimetry of BL Lac jets] 
{Probing dissipation mechanisms in BL Lac jets through X-ray polarimetry} 

\author[Tavecchio et al.]
{F. Tavecchio$^1$\thanks{E--mail: fabrizio.tavecchio@brera.inaf.it}, M. Landoni$^1$, L. Sironi$^2$ and P. Coppi$^3$
\\
$^1$INAF -- Osservatorio Astronomico di Brera, via E. Bianchi 46, I--23807 Merate, Italy\\
$^2$Department of Astronomy, Columbia University, 550 W 120th St, New York, NY 10027, USA\\
$^3$Yale Center for Astronomy and Astrophysics, Yale University, New Haven, CT 06520-8121, USA
}

\begin{document}

% \date{Accepted 1988 December 15. Received 1988 December 14; 
% in original form 1988 October 11}

%\pagerange{\pageref{firstpage}--\pageref{lastpage}} \pubyear{2007}

\maketitle

\begin{abstract} 
The dissipation of energy flux in blazar jets plays a key role in the acceleration of relativistic particles. Two possibilities are commonly considered for the dissipation processes, magnetic reconnection -- possibly triggered by instabilities in magnetically-dominated jets -- , or shocks -- for weakly magnetized flows. We consider the polarimetric features expected for the two scenarios analyzing the results of state-of-the-art simulations. For the magnetic reconnection scenario we conclude, using results from global relativistic MHD simulations, that the emission likely occurs in turbulent regions with unstructured magnetic fields, although the simulations do not allow us to draw firm conclusions. On the other hand, with local particle-in-cell simulations we show that, for shocks with a magnetic field geometry suitable for particle acceleration, the self-generated magnetic field at the shock front is predominantly orthogonal to the shock normal and becomes quasi-parallel downstream. Based on this result we develop a simplified model to calculate the frequency-dependent degree of polarization, assuming that high-energy particles are injected at the shock and cool downstream. We apply our results to HBLs, blazars with the maximum of their synchrotron output at UV-soft X-ray energies. While in the optical band the predicted degree of polarization is low, in the X-ray emission it can ideally reach 50\%, especially during active/flaring states. The comparison between measurements in the optical and in the X-ray band made during active states (feasible with the planned {\it IXPE} satellite) are expected to provide valuable constraints on the dissipation and acceleration processes.
\end{abstract}


\section{Introduction}

Extragalactic relativistic jets shine through the entire electromagnetic spectrum, from radio frequencies up to very-high energy gamma rays (e.g. Romero et al. 2017). The emission properties of jets are best studied in blazars (Urry \& Padovani 1995) for which -- thanks to the favorable geometry --  the relativistically boosted non-thermal radiation produced by the plasma flowing along the jet outshines any other radiative component associated to the active nucleus (accretion disk, broad lines, dust emission). The radiative bolometric output of blazars accounts for a sizable fraction ($\gtrsim 10$\%) of the inferred energy flux carried by the jet (e.g. Celotti \& Fabian 1993, Tavecchio et al. 2000, Celotti \& Ghisellini 2008) which, in turn, can be as large as the power brought onto the central black hole by the accreting material (e.g. Ghisellini et al. 2014).

Despite intensive investigation, the understanding of the basic processes leading to the dissipation of the jet energy flux (and the subsequent energization of the charged particles responsible for the emission) is far from clear. Current simulations converge to suggest two main general routes by which the energy carried by a relativistic jet can be dissipated and made available to be tapped by relativistic particles: {\it a)} for initially magnetically dominated jets, recent simulations support the view that a fraction of the magnetic energy can be dissipated through  reconnection, possibly triggered by the onset of non-linear (kink) jet instabilities (Begelman 1998). At reconnection sites particles can be efficiently accelerated forming non-thermal distributions, as clearly shown by particle-in-cell simulations (e.g. Zenitani \& Hoshino 2001, Sironi \& Spitkovsky 2014; Melzani et al. 2014; Guo et al. 2014, 2015; Sironi et al. 2016; Werner et al. 2016, Kagan et al. 2015); {\it b)} for weakly magnetized flows, instead, the most likely dissipation process involves a shock (e.g.  Aller, Aller, \& Hughes 1985, Kirk et al. 1998), responsible for the particle energization through the classical shock diffusive acceleration mechanism (Blandford \& Eichler 1987; Sironi et al. 2009, 2011, 2013, Spitkovsky 2008).

From the observational point of view, to make a distinction between the two alternative scenarios is not an easy task. Limits on the magnetization of the plasma associated to the emission region (and thus, indirect constraints to the reconnection scenario) can be obtained from the modeling of the spectral energy distributions (SED) of blazars with standard emission models (e.g. Nalewajko et al. 2014, Tavecchio \& Ghisellini 2016) but this method is not fully conclusive because of possible complications of the basic one-zone emission scheme (e.g. Sironi et al. 2015). In this context, polarimetry is considered a powerful tool to break the degeneracies involved in the SED modeling and to get insight into the magnetic field or emission region geometry (Angel \& Stockman 1980; see also Barres De Almeida et al. 2009, 2014). Optical polarimetry is particularly attractive for sources, such as the blazar subclass of BL Lacs, in which the radiation in the optical band is produced by synchrotron emission. Several studies focused on the optical polarimetric properties of BL Lacs consistently show a relatively low degree of polarization during low/quiescent states (floating around 5-10\%), slightly raising to 10-20\% during active states (e.g., Tommasi et al. 2001a,b, Larionov et al. 2013, Sorcia et al. 2013, Covino et al. 2016, Hovatta et al. 2016 among the most recent). This suggests a low degree of order for the underlying magnetic fields. 

Recently, the subject has been revitalized by observations revealing characteristic patterns displayed by the degree of polarization and the polarization angle in the optical band in correspondence of energetic flares (e.g. Marscher et al. 2008, 2010, Abdo et al. 2010, Larionov et al. 2013). These events could be  understood as the signature of an emission region moving down a jet with a substantial toroidal field (e.g. Marscher et al. 2008, Zhang et al. 2014, 2015). Alternatively, such patterns could be the result of strong turbulence in the flow. Indeed, Marscher (2014) showed that several properties of blazars can be accounted for by a model where the observed emission from blazars is the incoherent sum of the emission produced by several turbulent cells chaotically evolving downstream of a standing shock. At zeroth order, in this framework the observed emission does not carry any information on the global geometry of the magnetic field in the jet, since its properties are mainly determined by the turbulent nature of the flow, which also governs the properties of the magnetic fields.

In this work we intend to take a different approach. Inspired by  simulations built for the general scenarios for jet dissipation sketched above, we investigate whether the different environments envisaged in the two schemes (i.e. shocks in a weakly magnetized flow vs magnetically dominated unstable jet) could result in different properties for the polarization of the emitted radiation, that could be possibly exploited to distinguish between the two alternatives or, more generally, to probe the magnetic structure of the jet. Anticipating some of the following results, in scenario a) the evolution expected for the kink instability development leads to particle acceleration occurring beyond the point where instabilities are triggered. In these regions any possible originally well organized toroidal field is expected to have gained a substantial poloidal component. Therefore, since there is no preferred direction for the projection of the magnetic field onto the sky, we do not expect a high degree of polarization. For the case b) the situation is expected to be quite different. In fact, for the shock case -- for  magnetic field configurations which allow efficient particle acceleration -- we expect that the component of the field perpendicular to the shock axis dominates in the shock vicinity (as a result of self-generated magnetic waves), while far away from the shock the perpendicular component decays, and the dominant magnetic field is quasi-parallel to the axis. Since rapidly-cooling high-energy  electrons emit close to the shock, we expect a high degree of polarization for the high-energy synchrotron emission (falling in the X-ray band in the specific case of so-called high-peaked BL Lac objects). Low energy electrons are instead expected to fill more homogeneously the post-shock region (or even to accumulate far downstream), thus experiencing a more isotropic magnetic field. Their synchrotron emission (expected to fall in the optical-IR) is thus expected to be characterized by a small degree of polarization. Measurements of the degree of polarization taken simultaneously in the optical and in the X-ray band by the planned {\it IXPE} satellite (Weisskopf et al. 2016) 
%and the proposed {\it XIPE} [Soffitta et al. 2013] satellites) 
are then expected to provide us an effective probe to study the dissipation and acceleration mechanisms working in the jets. As said, the situation is particularly favorable considering the BL Lac objects whose synchrotron component peaks in the UV-soft X ray band (the so-called HBL), for which we can characterize the polarimetric properties both for high-energy, freshly accelerated and low-energy cooled electrons. 

The aim of this paper is to provide a first exploration of the polarimetric features that we can expect from the two alternative scenarios for the energy dissipation and to provide some testable predictions for the specific case of the shock acceleration scheme. Specifically, in \S 2 we discuss the two scenarios for dissipation, reporting some quantitative estimates on the expected geometry of the magnetic field in the emission region based on state-of-the-art numerical simulations. In \S 3 we present a toy model to calculate the frequency dependent polarization degree in the case of the shock scenario and we provide some results and predictions. In \S 4 we discuss our results.

\begin{figure*}
 \centering
 \hspace*{-0.0truecm}
 \includegraphics[width=0.9\textwidth]{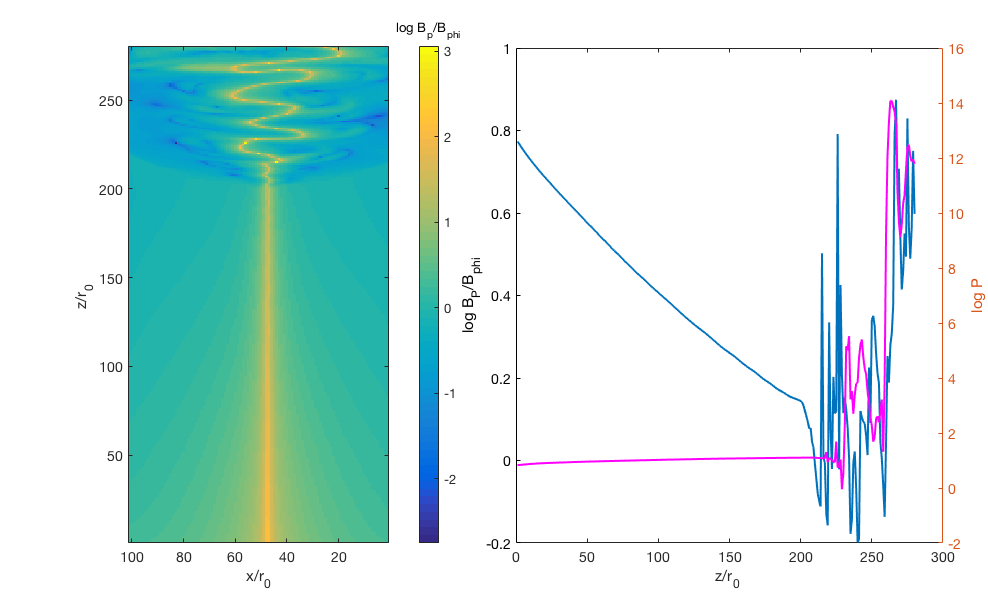}\\%{FiguraMappa.eps}\\
 \caption{Left: map of the logarithm of the ratio of poloidal to toroidal magnetic fields from the simulation of Barniol Duran et al. (2017) discussed in the text. One can clearly see the onset of a strong non-linear kink-like instability at $z/r_0\sim 200$. Right: the blue line shows the profile of the cross-section averaged $B_{p}/B_{\phi}$ ratio. The magenta line shows the profile of the pressure along the jet (in arbitrary units) derived as explained in the text.}
 \label{fig:rodolfo}
\end{figure*}

\section{Scenarios for dissipation and particle acceleration}

In the next two sections we analyze the results of two simulations of the two scenarios described above. Our aim is to gain some insights into the expected geometry of the magnetic field in the emission regions and, therefore, on the expected polarization properties.

\subsection{Kink-instability driven magnetic reconnection}

It is widely recognized that a jet dominated by a toroidal magnetic field ($B_{\phi}$) can easily develop  kink instabilities, potentially leading to the onset of dissipation through magnetic reconnection (e.g. Begelman 1998, Giannios \& Spruit 2006). On the other hand, jets showing a significant toroidal component are the natural outcome of simulations for jet formation (e.g. McKinney \& Narayan 2007, Bromberg \& Tchekhovskoy 2016). Therefore, it is natural to suppose that instabilities could provide a natural way to account for jet dissipation and possibly for the polarization properties of jets (e.g. Zhang et al. 2017, Nalewajko 2017).

Recently, Barniol Duran et al. (2017) used  state-of-the-art 3D MHD simulations of magnetically dominated jets (performed with the HARM code, Gammie, McKinney \& Toth 2003) to study the development of the kink instability and its effects on the jet propagation. They consider two cases, namely a jet which is excavating its way through an external medium and a jet which is propagating in an already evacuated channel. In the former case the simulations show the development of a strong kink mode, eventually leading to a substantial reduction of the toroidal field component. At the same time, the jet, interacting with the external medium, suffers a re-collimation resulting in the formation of a conical internal shock, responsible for a minor part of the observed energy flux dissipation. 

An important point, not explicitly discussed by Barniol Duran et al. (2017), is the correlation between  the maximum of the pressure inside the jet (resulting from the jet Poynting flux dissipation) and the magnetic field geometry. The simple idea is that if a large fraction of the dissipation (and, by extension, of non-thermal particle acceleration) happens to occur before a substantial reduction of the toroidal component has taken place, we could expect a high degree of polarization of the resulting radiation, determined by the mostly ordered toroidal field. We use the 2D slices for entropy ($s$) and density ($\rho$) extracted from the Barniol Duran et al. (2017) A2-3D simulation (kindly provided us by R. Barniol Duran; reported in their Fig. 9 and 11) to calculate the cross-section averaged profile (from $x/r_0=-30$ to $x/r_0=30$, see Fig. \ref{fig:rodolfo}; $r_0$ is a normalization corresponding to the injection distance of the jet from the compact object used in the simulation) of the pressure\footnote{derived from $s$ and $\rho$ using the expression for the entropy of adiabatic flows, $s=\ln(p/\rho^{\gamma_a})/(\gamma_a-1)$, using the relativistic value for the adiabatic index, $\gamma_a=4/3$.} (assumed to be a good proxy also for the energy density of the non-thermal particles; in other words, we assume that the non-thermal particle energy density is a fixed fraction of the particle energy density). The resulting averaged profile of the pressure along the jet is plotted in Fig.\ref{fig:rodolfo} (right panel), together with the corresponding averaged profile of $B_{\rm p}/B_{\phi}$ (blue line) and the map of $B_{\rm p}/B_{\phi}$ (left panel). The profile of  $B_{\rm p}/B_{\phi}$ shows the expected decreasing trend (dictated by $B_{\rm p}\propto z^{-2}$ and $B_{\phi}\propto z^{-1}$) until the instability sets in around $z/r_{0}\approx 200$. After that point the ratio displays huge variations, but with a clear increasing trend, marking the progressive reduction of the $B_{\phi}$ component. The pressure profile displays a qualitatively similar behavior. The pressure is very low until the onset of the instability (in fact the jet is initially injected as a cold flow), where it starts to increase, following the progressive dissipation of part of the magnetic jet energy. The maximum of the pressure clearly occurs beyond $z/r_{0}\approx 250$, in regions where the poloidal field is of the same order (or slightly larger) than the toroidal one.
From this analysis we tentatively conclude that the the bulk of the dissipation (and hence of the emission) occurs in regions in which the toroidal and the poloidal components are comparable and, therefore, we do not expect to detect high polarization (at any frequency). However, we remark that a firmer conclusion would require more refined simulations, with a resolution sufficient to better identify the sites of particle acceleration and characterize the corresponding magnetic field structure. This will be particularly important at high frequencies, where the cooling length might be comparable or smaller than the size of the acceleration region
Furthermore, a far better choice would be to gauge the efficiency of particle acceleration not on the local energy density but on the local instantaneous dissipation rate, as in the work by Zhang et al. (2017).\footnote{We would like to note that in Zhang et al. (2017)  the assumption about the existence of a fixed background of non-thermal particles could help to artificially boost the polarization.}

Finally it is important to mention that another possible framework based on magnetic reconnection, not considered here but worth to be investigated, is the one envisioning a so-called striped wind. This scenario has been considered for the emission from pulsar wind nebulae and jets (e.g. Spruit at al. 2001, Zrake 2016) and postulates the existence of a magnetic field configuration characterized by alternating magnetic toroidal fields, which provides an ideal geometry for reconnection.

\subsection{PIC simulation of a trans-relativistic magnetized shock}

In this section, we describe the results of a particle-in-cell (PIC) simulation of a magnetized shock. The analysis performed by Sironi et al. (2015) constrains the parameter space of shocks in blazars. For shocks to be efficient in powering the jet emission, the authors find that they need to be
at least mildly relativistic, i.e., with the relative Lorentz factor between upstream (pre-shock) and downstream (post-shock) regions of $\Gamma_{0}\gtrsim $ a few. To attain rough equipartition between electron and magnetic energy in the shock downstream, as inferred from modeling the spectral energy distribution of blazar sources,
the magnetization in the shock upstream has to be in the range $\sigma=B_0^2/(\Gamma_0-1) m_i n_{0,i} c^2 \sim 0.01-0.1$. Here, $m_i$ is the proton mass, $c$ is the speed of light, and $B_0$ and $n_{0,i}$ are the values of the magnetic field and of the proton density in the upstream region far from the shock, measured in the post-shock frame.
Under these conditions, for most magnetic field orientations, Sironi et al. (2015) find that shocks are superluminal (Begelman \& Kirk 1990). Here, charged particles are constrained to follow the field lines, whose orientation prohibits repeated crossings of the shock. So, the particles have no chance to undergo Fermi acceleration, which instead is required to explain the broadband non-thermal emission signatures of blazar jets. Sironi et al (2015) find that the shock model for blazar emission can work only in a small region
of the parameter space, with $\Gamma_0\sim 1.5$, $\sigma\sim 0.01-0.1$ and $\theta_{B,0}\lesssim 30^\circ$. Here, $\theta_{B,0}$ is the angle between the upstream magnetic field and the shock normal, as measured in the post-shock frame.

Here, we use the three-dimensional (3D) electromagnetic PIC code TRISTAN-MP (Spitkovsky 2005) to follow the evolution of a trans-relativistic ($\Gamma_0=1.5$) magnetized ($\sigma=0.1$) shock propagating into an electron-proton plasma (for computational convenience, the mass ratio is $m_i/m_e=25$). The simulation is performed in the post-shock (downstream) frame. The upstream plasma is initialized with a negligible thermal dispersion ($k_BT/m_i c^2=10^{-4}$) and the upstream magnetic field forms an angle $\theta_{B,0}=10^\circ$ with respect to the shock normal, as measured in the downstream frame of the simulation (i.e., the shock is ``quasi-parallel''). In the upstream medium, we also initialize a magnetic field $\bmath{B}_0$ (whose strength is parameterized with $\sigma$) and a motional electric field $\bmath{E}_0=-\mbox{\boldmath{$\beta$}}_0\times\bmath{B}_0$, where  $\mbox{\boldmath{$\beta$}}_0$ is the three-velocity of the injected plasma. The shock is set up by reflecting the upstream flow from a conducting wall located at $z= 0$, and it propagates along the $+\bmath{\hat{z}}$ direction. The simulation is performed on a 2D computational domain in the $yz$ plane (yet, all three components of particle velocities and electromagnetic fields are tracked), with periodic boundary conditions along $y$, whereas the $z$-extent of the domain increases with time at the speed of light (for a more detailed description of the computational setup, see Sironi \& Spitkovsky 2009, 2011, Sironi et al. 2013). Each computational cell is initialized with four particles (two per species), but we have tested that 
experiments with a larger number of particles per cell (up to 8 per species) are not changing our conclusions. The electron skin depth $c/\omega_{\rm pe}$ is resolved with 5 cells, where $\omega_{\rm pe}\equiv(4\pi e^2 n_{0,e} /\Gamma_0 m_e)^{1/2}$ is the relativistic  plasma frequency for the upstream electrons, with number density $n_{0,e}$ (measured in the simulation frame) and mass $m_e$. The simulation timestep is $\Delta t=0.09\,\omega_{\rm pe}^{-1}$. The extent of the box along $y$ is $\sim 46\comp$, where the proton skin depth is $\comp=\sqrt{m_i/m_e}c/\omega_{\rm pe}$. We have tried with 2D boxes whose $y$ extent is two times smaller, finding the same conclusions. The results that we present below refer to a pre-shock field lying in the $xz$ plane, but we have tested that our conclusions do not change if the initial field is lying in the $yz$  plane of the simulation.

\begin{figure}
 \centering
 \includegraphics[width=0.52\textwidth]{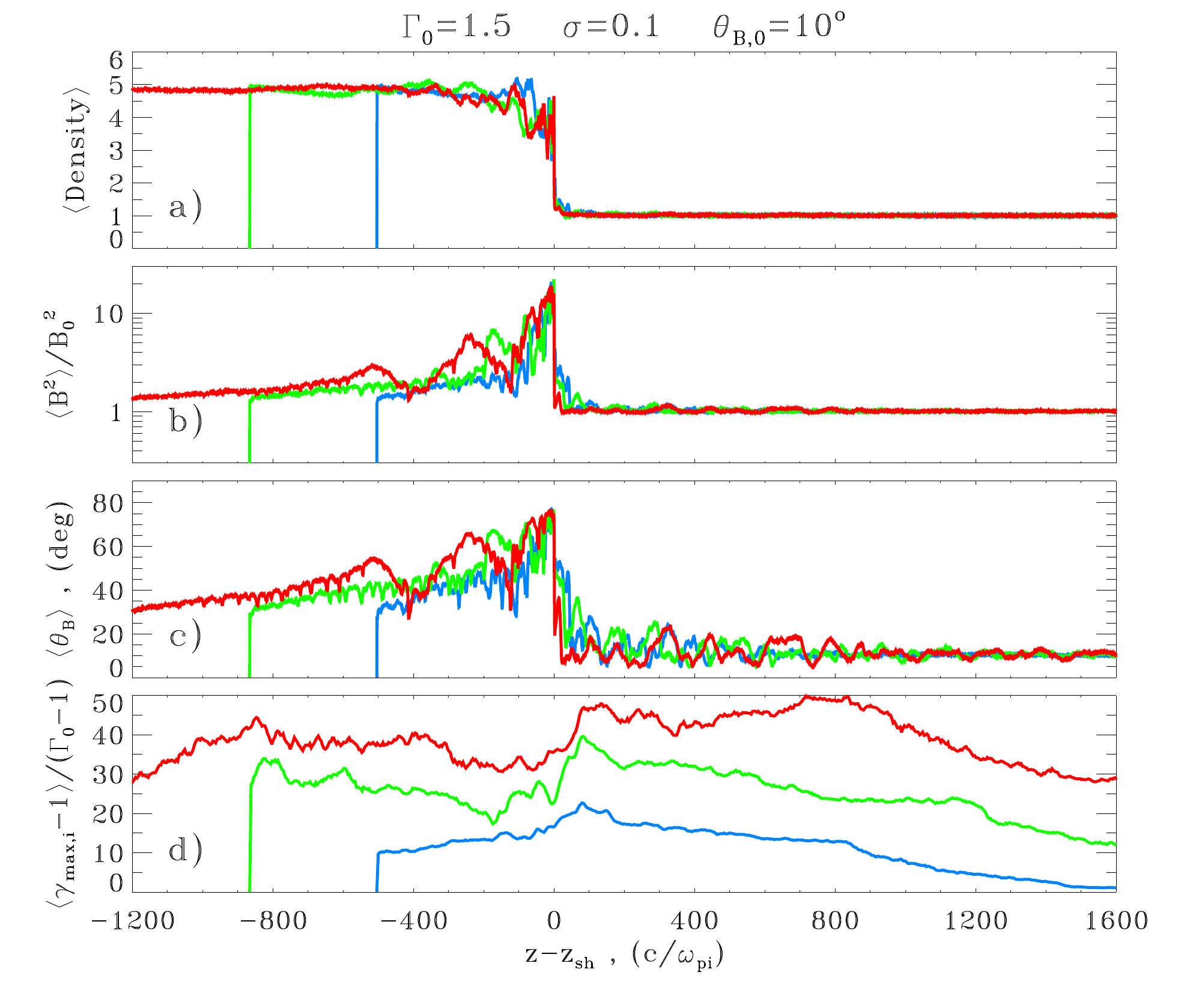}\\
 \caption{1D $y$-averaged profiles of various quantities, from a 2D PIC simulation of a trans-relativistic ($\Gamma_0=1.5$) magnetized ($\sigma=0.1$) quasi-parallel ($\theta_{B,0}=10^\circ$) shock propagating into an electron-proton plasma. The $z$ location is measured relative to the shock, in units of the proton skin depth $\comp$. At three different times ($\omega_{\rm pi}t\sim 2500$  in blue, $\omega_{\rm pi}t\sim 4300$ in green and $\omega_{\rm pi}t\sim 6100$ in red), we plot the following quantities: (a) particle number density, in units of the upstream value; (b) magnetic energy density, in units of the upstream value; (c) local obliquity angle of the magnetic field, relative to the shock normal; (d) maximum proton Lorentz factor, in units of the upstream bulk Lorentz factor.}
 \label{fig:shock}
\end{figure}

We follow the shock evolution over time, until $\omega_{\rm pi}t\sim 6100$. In \fig{shock}, we plot 1D profiles of various quantities, at three different times in the shock evolution: $\omega_{\rm pi}t\sim 2500$ (blue), $\omega_{\rm pi}t\sim 4300$ (green) and $\omega_{\rm pi}t\sim 6100$ (red). The horizontal axis shows the $z$-location measured with respect to the shock. The shock appears as a density compression by a factor of $\sim 5$ (panel (a)). While the magnetic energy (panel (b)) in the far downstream is consistent with the magnetohydrodynamic (MHD) compression of the upstream field, an excess of magnetic energy is observed at the shock and in the near downstream, up to a factor of $\sim 20$ higher than the upstream field energy. The excess magnetic energy at the shock comes from the compression of circularly-polarized  Alfv\'enic waves generated in the upstream region by shock-accelerated protons propagating ahead of the shock, in analogy to what has been observed in ultra-relativistic shocks by Sironi \& Spitkovsky (2011) and in non-relativistic shocks by Caprioli and Spitkovsky (2014). In fact, as shown by panel (d), shock-accelerated protons are present both upstream and downstream. There, we plot the spatial profile of the maximum proton Lorentz factor averaged over the $y$ direction. One can see that protons are accelerated to higher and higher energies over time (compare the blue, green and red lines in panel (d)), and that the population of high-energy protons extends over time further upstream and further downstream. The interaction of the upstream-propagating protons with the incoming flow is eventually responsible for the growth of Alfv\'enic waves in the upstream, whose compression at the shock drives the magnetic energy up to $\sim 20$ times the upstream value. As shown in panel (b), the decay length of the magnetic energy (i.e., the distance over which the magnetic energy goes back to the MHD expectation) is comparable to the $z$-extent of the downstream region, which increases linearly over time as the shock propagates to the right. 

The $y$-averaged profile of the obliquity angle of the magnetic field (panel (c)) traks the magnetic energy profile (panel (b)). Far downstream, where the magnetic field settles back to the MHD prediction, $\langle\theta_B\rangle\sim 30^\circ$, i.e., the field is not very inclined with respect to the shock direction of propagation. However, at the shock and in the near downstream,  $\langle\theta_B\rangle$ can be as large as $60^\circ-80^\circ$, i.e., the field is nearly perpendicular to the shock normal. As we discuss below, this will have important implications for the observed polarization signature.

\section{A toy model for polarization from shocks}

Inspired by the results of the numerical simulation discussed above, we implemented a simplified numerical model to quantify the polarization properties of a jet hosting a shock accelerating relativistic electrons. We limit ourselves to this case because the simulations provide a sufficiently clear framework. For the kink-instability triggered reconnection scenario, on the contrary, the information on the configuration of the dissipation region (particle acceleration site, geometry of the magnetic field) is less conclusive and we leave its treatment to a future work.

The basic ingredient of the model is the (rapidly decaying) magnetic field component perpendicular to the shock normal, generated by  instabilities in the vicinity of the shock front. This field is expected to imprint a well defined  polarization to the radiation emitted by electrons at the highest energies which, due to the strong radiative losses, can survive only very close to the shock. In typical HBL jets these electrons emit at X-ray energies. At lower frequencies (optical, IR), instead, the emission is produced by low energy electrons, which fill a good fraction of the the post-shock emitting region. In these conditions we expect that optically emitting electrons experience a (on average) less ordered field, resulting in a low degree of polarization, in agreement with the typically low polarization degree observed in the optical band.

\subsection{Setup}

\begin{figure*}
 \centering
 \hspace*{-0.0truecm}
 \includegraphics[width=1.05\textwidth]{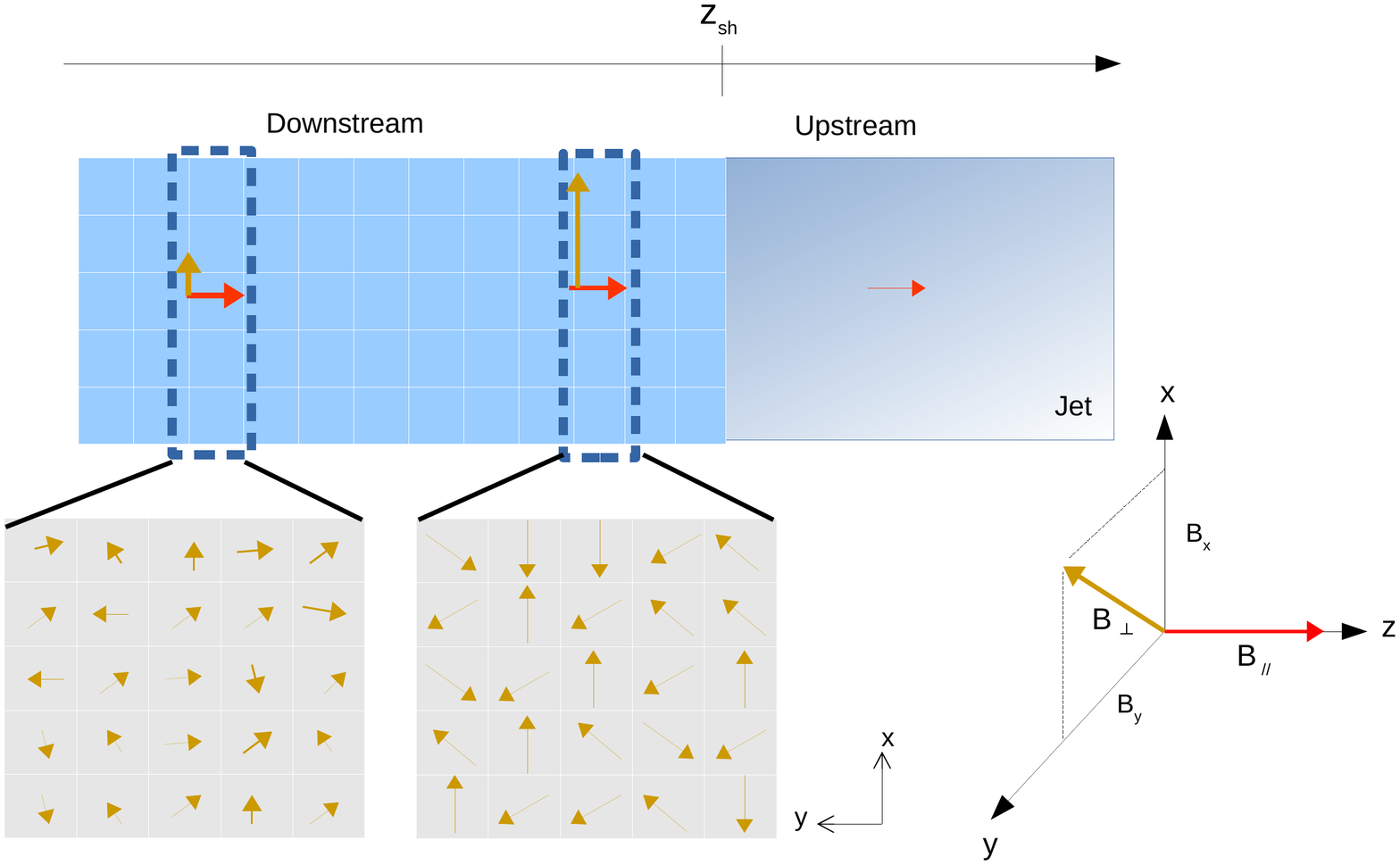}\\%{FiguraMappa.eps}\\
 \vspace*{-0.8truecm}
 \caption{Sketch of the setup used to calculate the polarization quantities in the shock scenario, drawn in the shock frame. The flow moves to the left along the $z$-axis, passing through the shock at $z_{\rm sh}$. In the upstream flow the magnetic field is predominantly parallel to the shock normal (red arrow). Downstream the magnetic field includes a normal component (yellow arrows), whose magnitude decreases with distance. The downstream flow is divided in cells (not to scale). In each cell the perpendicular component is assumed to have a random direction in the $x$-$y$ plane, as shown in the two lower panels reporting a cross section of the jet at two different distances from the shock (for simplicity we depict them as squares instead of circles). The observer line of sight in this frame lies in the $x$-$z$ plane.}
 \label{fig:cartoon}
\end{figure*}

We perform calculations of the frequency-dependent degree of polarization expected in the shock scenario by means of a quite simplified model in which we try to include all relevant physical processes (the geometry is sketched in Fig. \ref{fig:cartoon}). We assume that the shock front is perpendicular to the jet axis (in this geometry the field parallel to the shock normal is the poloidal field, while the perpendicular component is the toroidal one) and that it encompasses the entire jet cross section. The post-shock (downstream) reference frame is characterized by a Lorentz factor $\Gamma_{\rm d}$ with respect to the observer frame. The observer line of sight lies at an angle $\theta_{\rm v}\leq1/\Gamma_{\rm d}$ with respect to the jet axis. Note that for an angle $\theta_{\rm v}=1/\Gamma_{\rm d}$ the observer receives the photons emitted at an angle $\theta^{\prime}_{\rm v}=\pi/2$ in the downstream plasma frame. In this situation a perfectly perpendicular field would produce a maximally polarized synchrotron radiation with the electron vector polarization angle aligned along the jet axis. All physical quantities characterizing the system (with the exception of $\Gamma_{\rm d}$) are expressed in the downstream reference frame.

The expected geometry of the self-generated magnetic field is described assuming a cell structure, each cell intended to represent a coherence domain (see also the model by Marscher 2014). For simplicity, each domain is assumed to have a cylindrical shape (with the axis parallel to the jet axis), characterized by  radius $r$ and height $h=r$. In each cell/domain the total magnetic field $\mathbf{B}$ is uniform and we can describe it by its components $B_x$, $B_y$ (components of $\mathbf{B_{\perp}}$) and $B_z$ (which coincides with $\mathbf{B_{\parallel}}$, see Fig.\ref{fig:cartoon}).  For each domain, $B_z$ is fixed to the  value at the shock front. The decrease of the perpendicular component displayed by the simulations is modeled assuming that the maximum allowed value $B_{\perp,\rm max}$ decreases with distance from the shock front, $z$. For simplicity,  $B_{\perp,\rm max}$ is assumed to follow a power law profile, $B_{\perp,\rm max}(z)=B_{\perp, 0}(z/z_{\rm sh})^{-m}$. In each cell $B_{\perp}$ is then selected randomly sampling a flat probability distribution in the interval $(-B_{\perp,\rm max}, B_{\perp, \rm max})$. In turn, we select the $x$ and $y$ components assuming a flat probability distribution for the angle $\alpha= \tan^{-1} (B_y/B_x)$.

The relativistic particles accelerated at the shock front and injected into the flow, are supposed to be advected downstream with velocity $v_{\rm adv}\approx c/3$ (comoving, independent on the distance). Because of radiative losses, the energy of the electrons changes with time (we ignore any possible reacceleration due to scattering with turbulence) and thus with the distance from the shock. We assume that the injected electron energy distribution is a power law $N(\gamma)\propto \gamma^{-2}$, with $\gamma < \gamma_{\rm max,0}$. At a given distance  from the shock, $z$, the evolved distribution will have the same slope, but the maximum energy will change with distance, $\gamma_{\rm max}(z)$. Taking into account that the total averaged B-field is the sum of the constant parallel and the decreasing perpendicular components, $B(z)^2=B_{\parallel}^2+B_{\perp}^2(z)$, the maximum Lorentz factor of the distribution at each distance can be  derived by integrating the differential equation $mc^2 d\gamma/dt =-(4/3) \sigma_T c U_B \gamma^2$ with the substitution $t \to z/v_{\rm adv}$ and assuming $U_B=B(z)^2/8\pi$.
  
Knowing the magnetic field components for each domain, we derive the frequency dependent Stokes parameters for synchrotron emission in the {\it observer} frame, $U_{\nu,i}$, $Q_{\nu,i}$ and $I_{\nu,i}$, adopting the formulation of Lyutikov et al. (2005). The total polarization degree and the electron vector position angle  $\Pi_{\nu}$ are finally derived from the total Stokes parameters $U_{\nu}=\sum U_{\nu,i}$,  $Q_{\nu}=\sum Q_{\nu,i}$ and $I_{\nu}=\sum I_{\nu,i}$, by using the standard formulae $\Pi_{\nu} = \sqrt{Q_{\nu}^2+U_{\nu}^2}/I_{\nu}$ and
\begin{equation}
\cos 2\chi_{\nu}=\frac{Q_{\nu}}{\sqrt{Q_{\nu}^2+U_{\nu}^2}}, \;\; \sin 2\chi_{\nu}=\frac{U_{\nu}}{\sqrt{Q_{\nu}^2+U_{\nu}^2}}.
\end{equation}

\subsubsection{Numerical implementation}

The computation of polarisation degree, from the Stokes parameters of each cell is a computationally-heavy task.
In order to implement a versatile tool that could be exploited to setup different scenarios and models, we
coded a {\tt C++} program divided in two parts. The first one is in charge to produce the the physical parameters of each cell while the second incorporates them to calculate, for each cell and each
frequency, the associated Stokes parameters. The overall output is then stored into a large no-SQL database (various GB per each model) that is is analysed using a Map-Reduce query (Dean \& Ghemawat 2004) to obtain the integrated Stokes parameters for each frequency. We exploited the Google Cloud Platform based BigData suite to perform the MapReduce task to minimize the costs and increase as much as
possible the efficiency in terms of computational time.

\subsection{Results}

\begin{figure}
 %\centering
 \hspace*{-0.8truecm}
 \vspace*{-0.4truecm}
 \includegraphics[width=0.55\textwidth]{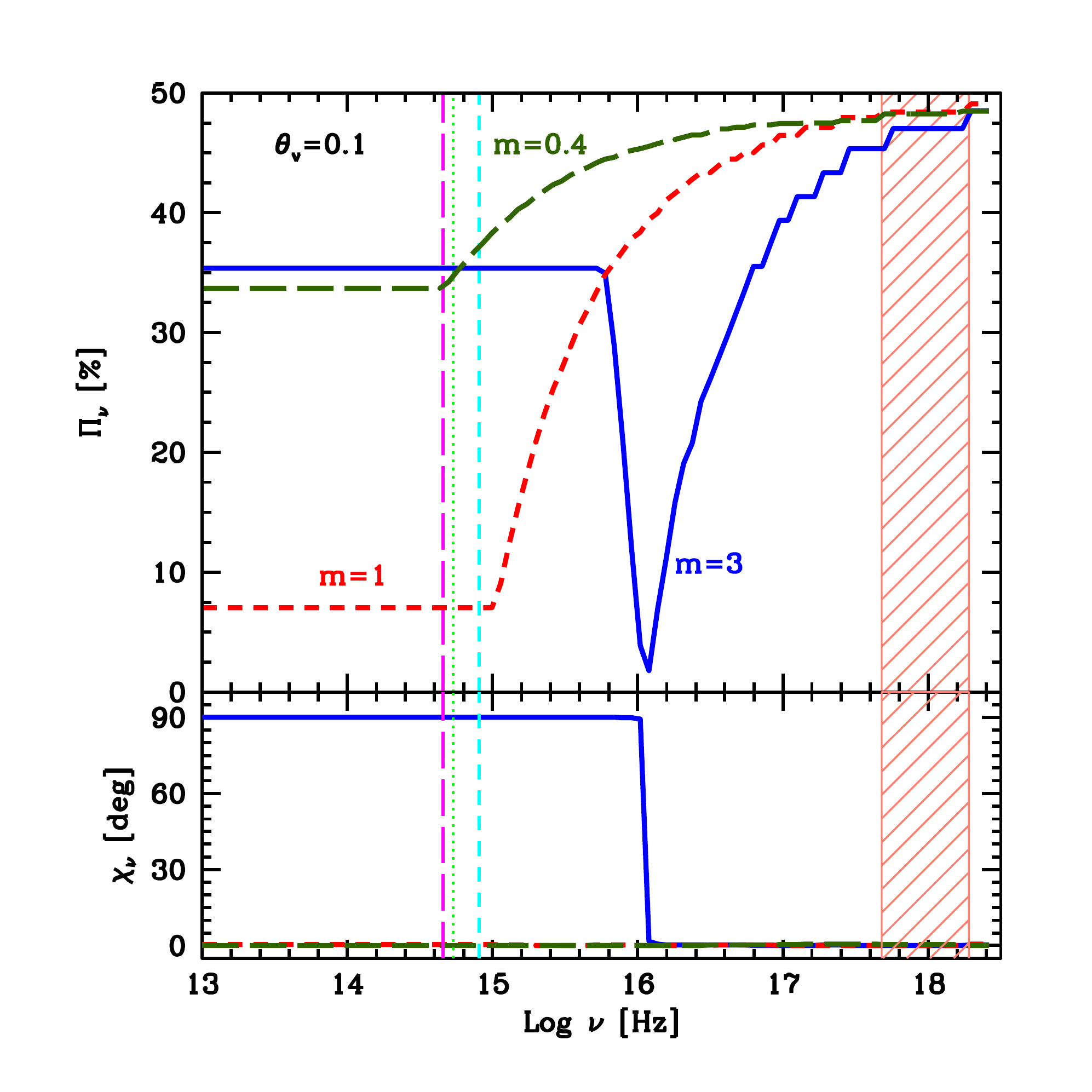}\\
 %\vspace*{-2.3truecm}
 \caption{Top panel: degree of polarization as a function of the observed frequency, $\Pi_{\nu}$, calculated for the shock scenario with the model described in the text. The observing angle is fixed to $\theta_{\rm v}=0.1$ rad. The blue solid, red short dashed and green long dashed curves correspond to the case $m=3, 1$ and 0.4, respectively.  The orange dashed area marks the energy range (2-8 keV) probed by the X-polarimeter of the planned {\it IXPE}. The magenta (long dashed), green (dotted) and cyan (dashed) thin vertical lines report the frequencies corresponding to R, V and U optical filters, correspondingly. Bottom panel: electric vector polarization angle $\chi_{\nu}$ as a function of frequency for the cases reported in the top panel.}
 \label{fig:pi}
\end{figure}

In Fig. \ref{fig:pi} we show some examples of the profile of the degree of polarization $\Pi_{\nu}$ and polarization angle $\chi_{\nu}$ as a function of frequency (from optical up to X-ray band) calculated with the model described above for a viewing angle $\theta_{\rm v}=0.1$.

The modeling of the BL Lac SED, especially for those belonging to the HBL subclass (e.g. Tavecchio et al. 2010), suggests low value for the magnetic field  (e.g. Tavecchio et al. 2010, Tavecchio \& Ghisellini 2016). We thus fix $B_{z}=0.05$ G. We also fix $\gamma_{\rm max,0}=3\times 10^6$ and we assume $\Gamma_{\rm d}=10$ in all models.

\begin{figure}
 %\centering
 \hspace*{-0.9truecm}
 \vspace*{-0.3truecm}
 \includegraphics[width=0.55\textwidth]{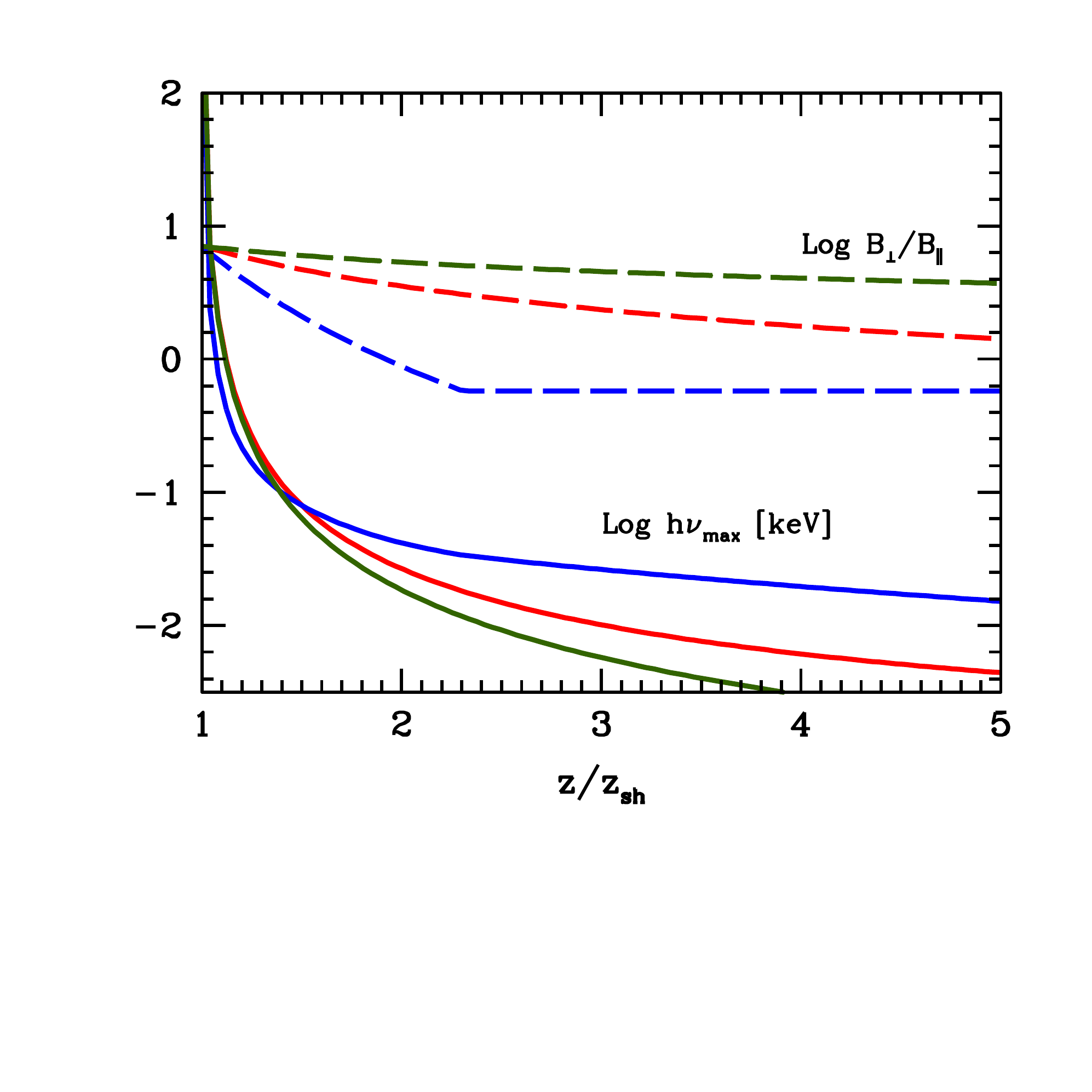}\\
 \vspace*{-2.3truecm}
 \caption{The solid curves report the logarithm of the ratio between the perpendicular and the parallel magnetic field component as a function of the distance from the shock front for the physical setup used to produce the curves in Fig. \ref{fig:pi}. The profiles correspond to $m=3$ (blue) and $m=1$ (red) and $m=0.4$ (green). The solid curves show the  maximum synchrotron photon energy (in the observer frame) as a function of the distance from the shock front. The curves correspond to the set of parameters given in the text. It is evident that the radiation in the X-ray range ($h\nu_{\rm max}\gtrsim1$ keV) is  produced in the region where $B_{\perp}\gg B_{\parallel}$.}
 \label{fig:param}
\end{figure}

The intensity of the perpendicular component of the magnetic field is assumed to decrease with the distance from the shock with a power law profile with $m=3$ (solid blue line), $m=1$ (dashed red line) and  $m=0.4$ (long dashed green line). We fix the value of $B_{\perp}$ at the shock, assuming a ratio for the energy density of the perpendicular and parallel components $B_{\perp}^2/B_{\parallel}^2=50$. This implies that the initial total magnetic field is $B_0=0.35$ G. We also assume, as for the simulations in Sect. 2.2, an inclination of the upstream magnetic field lines with respect to the shock normal (measured in the downstream frame) of about 10 degrees. The MHD jump conditions prescribe that the inclination of the field lines in the post-shock region is about 30 degrees. This condition holds at some distance from the shock front, in regions where the self-generated field has already been dissipated. Therefore, for the cases in which the perpendicular component, following the power law profile described above, reaches at some distance the value fixed by the jump conditions, $B_{\perp, {\rm MHD}}$, we assume a constant value $B_{\perp}=B_{\perp,{\rm MHD}}$ beyond that distance.
We assume a jet radius $r=10^{15}$ cm and a downstream region extending from $z=z_{\rm sh}=10^{16}$ cm to $z=5\times 10^{16}$ cm. The profiles of the ratio between the perpendicular and the parallel component of the magnetic field and those of the maximum synchrotron frequency are shown in Fig. \ref{fig:param}.

%Furthermore we calculated $\Pi$ for three values of the jet viewing angle, i.e. $\theta_{\rm v}=0.1$    (corresponding to $1/\Gamma_{\rm d}$), $0.05$ and 0.025 rad.

For this benchmark set of parameters -- suitable to describe HBL jets -- the electrons at $\gamma_{\rm max,0}$ emit their synchrotron radiation at an observed frequency $\nu_{\rm max}\approx 3\times 10^6 \gamma_{\rm max,0}^2 B_0^2 \Gamma \; {\rm Hz}= 10^{20}$ Hz, about 300 keV (see Fig. \ref{fig:param}). However, for these electrons the cooling length is extremely short, $z_{\rm cool} = v_{\rm adv} t_{\rm cool}\approx 2\times 10^{13}$ cm (assuming $v_{\rm adv}\sim c/3$). For electrons emitting in the keV range, with Lorentz factors approximately 20 times lower, i.e. around $10^5$, the corresponding cooling length is about 20 times larger, i.e. few $10^{14}$ cm, still short compared to the distance required for the substantial reduction of the perpendicular component of the magnetic field (dashed lines in Fig. \ref{fig:param}). The net effect of this rapid cooling is that the emission in the X-ray band mainly occurs in regions with $B_z\ll B_{\perp}$. For the case under consideration here, with $\theta_{\rm v}=1/\Gamma_{\rm d}=0.1$, the observer detects the radiation as if he is observing in the jet rest frame in a direction orthogonal to the jet axis. In this case, in the X-ray band the observer measures a very high polarization degree, $\Pi \approx 50$\% since, by construction, the magnetic field projected in the sky has a dominant direction (orthogonal to the jet axis). The corresponding electric vector is oriented along the jet axis, implying $\chi=0^{\circ}$ (bottom panel in Fig. \ref{fig:pi}).

Electrons emitting in the optical range, on the other hand, are characterized by long cooling timescale, and can occupy the entire post-shock volume. Therefore they will also emit in regions characterized by a lower perpendicular field. At these frequencies the actual observed degree of polarization will therefore represent an average over the entire volume. In particular one can expect that the  optical polarization measured by the observer is lower than the value observed at X-ray energies. In reality, the situation is a bit more complex. Let us focus on the case $m=3$ (blue solid line in Fig. \ref{fig:pi}). In this case the perpendicular field decreases quite fast and reaches the MHD value at distances $z\sim 2 z_{\rm sh}$ (see Fig. \ref{fig:param}). This implies that the emission at frequencies in the optical band occurs mainly in regions in which the magnetic field is dominated by the parallel component. In this case we expect a transition, from frequencies mainly produced in regions with dominant (but decreasing) {\it perpendicular} field to frequencies mainly produced in regions dominated by the {\it parallel} component. This is the reason of the behavior displayed by the solid lines in Fig. \ref{fig:pi}: while $\Pi$ decreases smoothly from hard to soft X-ray energies, it displays a sudden jump to $\Pi\approx 35\%$ in the far UV band, accompanied by the rotation by $\Delta \chi =90^{\circ}$ of the polarization plane (bottom panel). For less rapid decrease of the perpendicular component ($m=1$, red dashed line), the ratio $B_z/B_{\perp}$ is less than 1 in the entire emitting volume (Fig. \ref{fig:param}) and this effect is not visible. The degree of polarization in the optical band is quite in agreement with the value usually measured in BL Lac, $\Pi\sim 10\%$. For even lower value of the slope, ($m=0.4$, green long-dashed line), the average value of the perpendicular component remains high for the entire emitting volume, determining  a large degree of polarization, $\Pi\sim 30\%$.

\begin{figure}
 %\centering
 \hspace*{-0.8truecm}
 \vspace*{-0.4truecm}
 \includegraphics[width=0.55\textwidth]{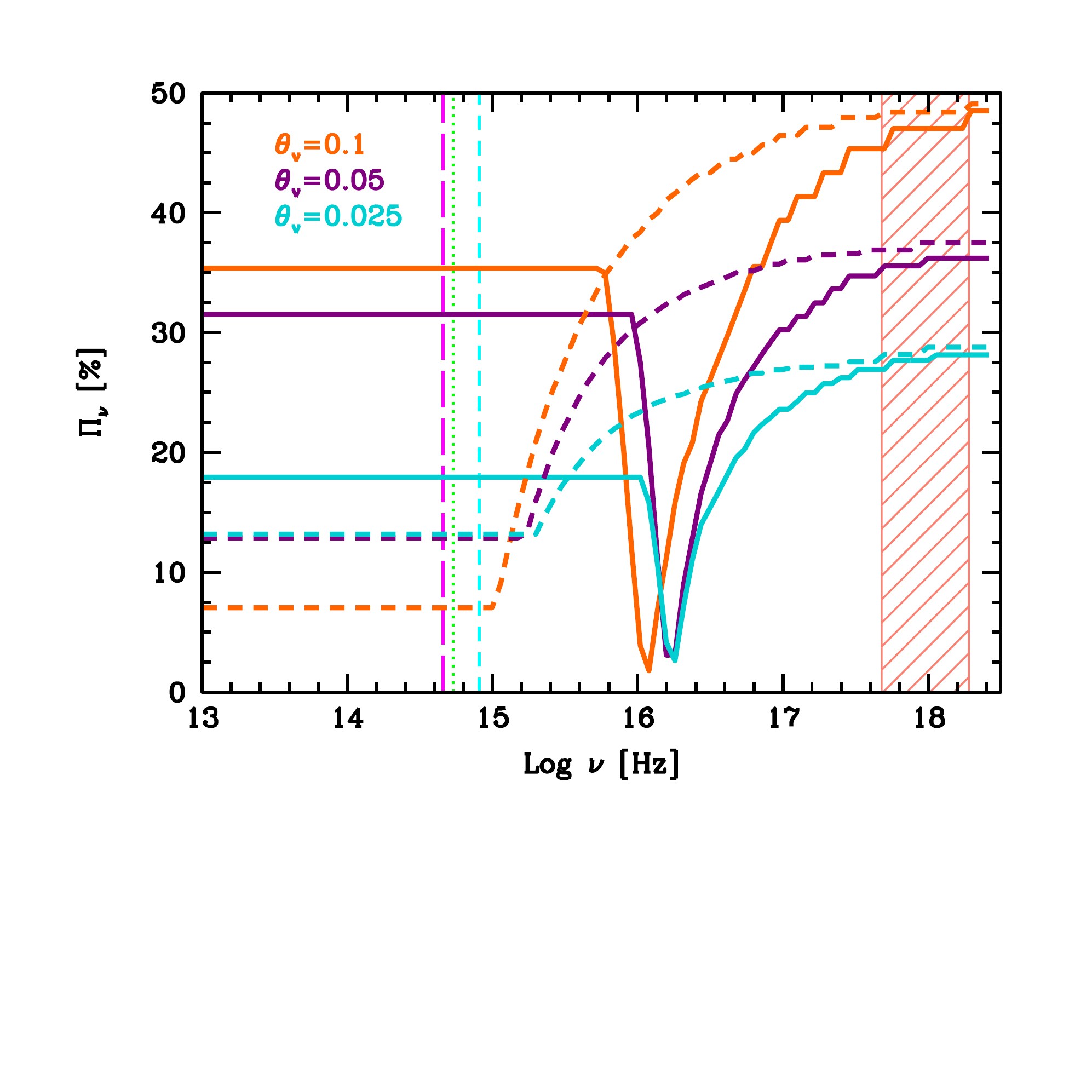}\\
 \vspace*{-2.3truecm}
 \caption{Degree of polarization as a function of the observed frequency, $\Pi_{\nu}$, calculated for different viewing angles, $\theta_{\rm v}=0.1$ (orange), 0.05 (violet) and 0.025 (cyan) radians. The solid (dashed) curves correspond to the case $m=3$ ($m=1$) and observing angles of $\theta_{\rm v}=0.1$ (red), 0.05 (blue) and 0.025 (green). The orange dashed area marks the energy range (2-8 keV) probed by the X-polarimeter of the planned {\it IXPE}. The magenta (long dashed), green (dotted) and cyan (dashed) thin vertical lines report the frequencies corresponding to R, V and U optical filters, respectively.}
 \label{fig:pimulti}
\end{figure}

The dependence of the degree of polarization with the viewing angles is illustrated by Fig. \ref{fig:pimulti}. We note that for angles smaller than $\theta _{\rm v}=1/\Gamma_{\rm d}$ (which implies angles in the jet frame lower than 90 deg), the magnetic field projected in the sky appears to have a component parallel to the jet axis (arising from the the projected perpedicular component) even if $B_{z}=0$. This determines a maximum value of the degree of polarization, again achieved at X-ray frequencies, less than the value found for  $\theta_{\rm v}=0.1$.  In particular, we found $\Pi\simeq 35-40$\% for $\theta_{\rm v}=0.05$ rad and $\Pi\simeq 30$\% for $\theta_{\rm v}=0.025$ rad. The degree of polarization in the optical band is affected by similar geometrical projection effects. For instance, for the case with $m=3$ the projected $B_z$ component is lower at smaller angles, therefore implying a reduction of the overall level of the polarization in this band that, as explained above, is sensitive to the average value of the $B_{\parallel}/ B_{\perp}$  ratio in the entire emitting volume. 

\section{Discussion}

Polarimetry is a powerful tool to investigate magnetic field geometries associated to astrophysical plasmas. In this work we have investigated the possible polarimetric signatures characterizing the dissipation sites in relativistic jets associated to blazars. We have focused in particular on the framework in which particle acceleration happens at shocks since current simulations offer a relatively definite scenario for the structure of the magnetic field in the post shock flow. 

We remark that our model is based on a very specific magnetic field geometry (i.e. quasi-parallel upstream magnetic field), since only in these conditions shocks are expected to efficiently accelerate particles (e.g. Sironi et al. 2015). Note also that, while the efficiency of electron acceleration is well characterized in ultra-relativistic shocks (Sironi \& Spitkovsky 2009, 2011, Sironi et al. 2013), where non-thermal electrons can account for up to 5-10\% of the post-shock energy, little is known on the corresponding efficiency in the trans-relativistic regime. However, given that non-relativistic shocks are generally inefficient electron accelerators (Park, Caprioli, Spitkovsky 2015), one expects that the efficiency in the trans-relativistic regime might be much smaller than in the ultra-relativistic case. If this is the case, the shock model might not satisfy the required energetics of the blazar emission (Sironi et al. 2015). Another caveat to our treatment is that we based our model on simulations able to reproduce only very small regions, much smaller then those involved in the blazar emission. In this sense, the simulations lacks a large-scale model of the structure embedding the shock and a self-consistent description of the orientation of the  the shock front relative to the jet axis. For simplicity we assume a shock whose front is perpendicular to the jet axis, but this could not be the case if, for instance, the shock is the result of re-collimation of the flow (e.g. Komissarov \& Falle 1997, Marscher 2014).

Our treatment remarks the importance of the organized structure of the magnetic field expected to be associated to shock (supported by simulations) in determining the polarimetric properties of the emission from blazars. In this sense our model could be considered complementary to the scenario discussed by Marscher (2014) in which, instead, the most relevant physical ingredient is the turbulence, expected to erase (or, at least, greatly smooth) any intrinsic geometry of the magnetic field. In particular, the dependence of both the magnetic field and the energy of the emitting particles on the shock distance naturally expected in our scheme provide a well defined trend of the degree of polarization with the frequency, a signature that is expected to be absent in the turbulent scenario.

In our work we focus on the polarimetric properties of the synchrotron radiation in the optical-X-ray range. One could wonder whether similar results apply to blazars in which the synchrotron component does not reach the X-ray band but peaks in the IR-optical band (the so-called Low peaked BL Lac, LBL). Electrons emitting in these sources in the optical band are close to the high-energy end of the energy distribution and should thus correspond to the X-ray emitting ones in HBL. If, as for X-ray emitting electrons in HBL, the cooling suffered by these electrons is sufficiently severe, they should exist only quite close to the shock and therefore one could expect a large polarization for their optical synchrotron radiation. However, a simple calculation shows that, assuming the physical parameters derived for these sources (e.g. Tavecchio et al. 2010) the cooling length for these electrons (which are characterized by Lorentz factors $\gamma\approx 10^4$) is of the order of $10^{16}$ cm, that is comparable to the extension of the post shock region. Therefore, for LBL even the electrons at the highest energies are expected to fill a large region of the downstream flow, determining a low degree of polarization. 
However, observationally LBL display, similarly to FSRQ, a degree of polarization larger than that of HBL. A possible reason for this is the presence of a globally order (helical) magnetic field carried by the jet (e.g. Lyutikov et al. 2005).

An important aspect that we have not considered in our preliminary inspection is variability, since we limited our study to the investigation of the polarimetric properties  of fixed shock structure. Indeed current observational (e.g. Larionov et al. 2013, Casadio et al. 2010, Kiehlmann et al. 2016, Bhatta et al. 2016, Carnero et al. 2017 among the most recent papers) and interpretative (Marscher et al. 2008, Abdo et al. 2010, Marscher 2014, Zhang et al. 2014, 2015) literature highlights the importance of variability in the study of polarimetric parameters. In a simple framework, if the energy flux of the incoming flow is modulated we expect variations of the dissipated energy, with associated variations of the emitted flux and of the shock structure. A proper investigation of these issues should be based on time-dependent simulations of the shock evolution. In any case we argue that our results are especially relevant when the emission of the jet is dominated by a unique active component, whose emission dominates over other simultaneously active regions.  These phases should correspond to flaring states, during which the good  correlation of the multifrequency light-curves support the idea that the radiation originates within a limited, compact emission region (e.g. Ulrich et al. 1997). 
This is also consistent with the fact that the measured degree of polarization in optical during flares is comparable with our estimates (20-30\%).
Low/quiescent states should be instead characterized by the contribution by several components and in that case the polarized flux could be ``diluted" by other less polarized radiation fields.

The targets suitable to investigate the issues discussed above are thus blazars emitting synchrotron radiation up to the medium-hard X-ray band during active states. As already mentioned, the sources belonging to the highly peaked BL Lacs are characterized by having the peak of their SED in the X-ray band. Ideal sources are thus the two brightest HBL in the sky, Mkn 421 and Mkn 501, widely monitored at all wavelengths, up to the TeV band. We used the web calculator PIMMS\footnote{\tt https://wwwastro.msfc.nasa.gov/ixpe/for\_scientists/pimms/} to derive the expected minimum degree of polarization (MDP) measurable at 99\% confidence level by {\it IXPE}. We adopt spectral parameters representative of Mkn 421 and Mkn 501 in their average-low state, namely a flux in the $2-10$ keV band $F_{2-10}=10^{-10}$ erg cm$^{-2}$ s$^{-1}$ and a photon index of 2.5. In these conditions the MDP is 12.8\% for 10 ksec of exposure. 
Increasing the flux by a factor of three , i.e. $F_{2-10}=3\times10^{-10}$ erg cm$^{-2}$ s$^{-1}$, with 10 ksec the reachable MDP is 8\%. Ideally, in such circumstances {\it IXPE} can detect the expected $\approx 30$\% polarization in less than 1 ksec, enabling to fully track the  evolution of the polarization during a flare. 

We conclude that polarization at the level envisaged by our scenario can be easily explored by upcoming instruments. In particular, observations during flares would provide key information on the evolution of the polarization and, in turn, on the underlying physical processes.

\section*{Acknowledgments}
We thank the referee for comments and criticisms. This work has been partly funded by PRIN-INAF 2014, ASI-INAF 2015-034.R.0 and PRIN-INAF SKA-CTA. LS acknowledges support from DoE DE-SC0016542, NASA Fermi NNX-16AR75G, NASA ATP NNX-17AG21G, NSF ACI-1657507, and NSF AST- 1716567. The  PIC simulations were performed on Habanero at Columbia. The authors thank HR for the precious help during the preparation of the paper. We are grateful to Rodolfo Barniol Duran for sharing the data of his simulations and to Dimitrios Giannios for discussions. FT and ML thank Yale and Columbia Universities for  hospitality during the preparation of this work. We are grateful to A. Bianco for providing partial funding for this research.

\end{document}